\documentstyle[12pt,epsfig,psfrag,wrapfig]{article}
\newcommand{\be}{\begin{equation}}
\newcommand{\ee}{\end{equation}}
\newcommand{\ba}{\begin{eqnarray}}
\newcommand{\ea}{\end{eqnarray}}
\newcommand{\Pnperp}{{\bf P}_{\!\!\!\perp\Nu}}
\newcommand{\Phperp}{{\bf P}_{\!\!\!\perp h}}
\newcommand{\Mn}{M_{\mbox{\tiny N}}}
\newcommand{\Nu}{  {\mbox{\tiny N}}}
\newcommand{\di}{ {\rm d} }

\newcommand{\simgeq}{\renewcommand{\arraystretch}{0.4}
		     \begin{array}{c} > \\ \sim \end{array}}
\newcommand{\la}{\langle}
\newcommand{\ra}{\rangle}
\newcommand{\athree}[3]{$\renewcommand{\arraystretch}{0.9}
	\begin{array}{c} \mbox{#1} \\  \mbox{#2} \\ \mbox{#3} \end{array}$}
\begin{document}
\title{\Large Predictions for azimuthal asymmetries in pion \& kaon 
       production in SIDIS off a longitudinally polarized deuterium 
       target at HERMES}
\author{\vspace{0.2cm}
	{\large 
	A.~V.~Efremov$^a$\thanks{Supported by RFBR Grant No. 00-02-16696 and 
			INTAS Project 587 and Heisenberg-Landau program.}, 
	K.~Goeke$^b$, 
	P.~Schweitzer$^{c}$} \\
	\vspace{-0.2cm}
	\footnotesize\it $^a$ Joint Institute for Nuclear Research, 
			 Dubna, 141980 Russia\\	
	\vspace{-0.2cm}
	\footnotesize\it $^b$ Institut f\"ur Theoretische Physik II, 
			 Ruhr-Universit\"at Bochum, Germany\\
	\vspace{-0.2cm}
	\footnotesize\it $^c$ Dipartimento di Fisica Nucleare e Teorica, 
			 Universit\'a degli Studi di Pavia, Pavia, Italy}
\date{}
\maketitle
\vspace{-7.5cm}
\begin{flushright}
Eur.~Phys.~J.~C (2002)\\
RUB/TP2-09/01
\end{flushright}
\vspace{5.5cm}
\begin{abstract}
\noindent
Predictions are made for azimuthal asymmetries in pion and kaon production 
from SIDIS off a {\sl longitudinally polarized deuterium} target for HERMES 
kinematics, based on information on the 'Collins fragmentation function' 
from DELPHI data and on predictions for the transversity distribution
function from non-perturbative calculations in the chiral quark-soliton 
model. There are no free parameters in the approach, which has been 
already successfully applied to explain the azimuthal asymmetries from 
SIDIS off polarized {\sl proton} targets observed by HERMES and SMC.
\end{abstract}
\section{Introduction}

Recently noticeable azimuthal asymmetries have been observed by HERMES in 
pion electro-production in semi inclusive deep-inelastic scattering (SIDIS) 
of an unpolarized lepton beam off a longitudinally polarized proton target 
\cite{hermes-pi0,hermes}. 
Azimuthal asymmetries were also observed in SIDIS off transversely 
polarized protons at SMC \cite{bravardis99}.
These asymmetries are due to the so called {\sl Collins effect}
\cite{collins} and contain information on $h_1^a(x)$ and $H_1^{\perp}(z)$.
The transversity distribution function $h_1^a(x)$ describes the 
distribution of transversely polarized quarks of flavour $a$ 
in the nucleon \cite{transversity}. 
The T-odd fragmentation function $H_1^{\perp}(z)$ describes the 
fragmentation of transversely polarized quarks of flavour $a$ 
into a hadron \cite{collins,muldt,Mulders:1996dh,hand}. 
Both, $H_1^{\perp}(z)$ and $h_1^a(x)$, are twist-2 and chirally odd. 
First experimental information to $H_1^{\perp}(z)$ has been extracted
from DELPHI data on $e^+e^-$ annihilation \cite{todd,czjp99}. 
HERMES and SMC data \cite{hermes-pi0,hermes,bravardis99} 
provide first information on $h_1^a(x)$ 
(and further information on $H_1^{\perp}(z)$). 

In ref.\cite{Efremov:2001cz} HERMES and SMC data 
on azimuthal asymmetries from SIDIS off a longitudinally and 
transversely, respectively, polarized {\sl proton} target 
\cite{hermes-pi0,hermes,bravardis99} have been well explained.
In the approach of ref.\cite{Efremov:2001cz} there are no free parameters:
for $H_1^{\perp}$ information from DELPHI \cite{todd,czjp99} was used,
for $h_1^a(x)$ predictions from the chiral quark-soliton model
were taken \cite{h1-model}.
In this note we apply this approach to predict azimuthal asymmetries 
in pion and kaon production from SIDIS off {\sl a longitudinally 
polarized deuterium} target, which are under current study at HERMES.

Similar works have been done in 
refs.\cite{Korotkov:1999jx,Anselmino:2000mb,Ma:2001ie} however making use 
of certain assumptions on $H_1^{\perp}$ and $h_1^a$ or/and 
considering only twist 2 contributions for transversal
with respect to the virtual photon momentum component of target
polarization. We take into account all $1/Q$ contributions.

\section{Ingredients for prediction: 
	 \boldmath $H_1^\perp$ and $h_1^a(x)$}

\paragraph{The T-odd fragmentation function \boldmath{$H_1^\perp$}.}
The fragmentation function $H_1^\perp(z,{\bf k}^2_\perp)$
describes a left--right asymmetry in the fragmentation of a transversely 
polarized quark with spin {\boldmath$\sigma$} and momentum  
${\bf k}$ into a hadron with momentum ${\bf P}_{\!h}=-z{\bf k}$. 
The relevant structure is 
 $H_1^\perp(z,{\bf k}^2_\perp)\,
  \mbox{\boldmath$\sigma$}({\mathbf k}\times
 {\mathbf P}_{\!\!\!\perp h}) / |{\bf k}|\la P_{\!\!\perp h}\ra$,
where $\la P_{\!\!\perp h}\ra$ is the average transverse momentum 
of the final hadron. Note the different normalization factor compared to
refs.\cite{muldt,Mulders:1996dh}: $\la P_{h\perp}\ra$ instead of $M_h$.
This normalization is of advantage for studying $H_1^\perp$ in chiral limit.

$H_1^\perp$ is responsible for a specific azimuthal asymmetry of 
a hadron in a jet around the axis in direction of the second hadron 
in the opposite jet. This asymmetry was measured using 
the DELPHI data collection \cite{todd,czjp99}.  
For the leading particles in each jet of two-jet events,
summed over $z$ and averaged over quark flavours 
(assuming $H_1^{\perp}=\sum_h H_1^{\perp\, q/h}$ is flavour independent), 
the most reliable value of the analyzing power is given by $(6.3\pm 2.0)\%$,
however a larger ``optimistic'' value is not excluded
\be\label{apower}
 	\left|{\la H_1^{\perp}\ra\over\la D_1\ra}\right| =(12.5\pm 1.4)\% 
	\ee
with presumably large systematic errors. 
The result eq.(\ref{apower}) refers to the scale $M_Z^2$ and to an
average $z$ of $\la z\ra\simeq 0.4$ \cite{todd,czjp99}.
A close value was also obtained from the pion asymmetry in 
inclusive $pp$-scattering \cite{Boglione:1999pz}.

%
%
%
	\begin{wrapfigure}{RD}{6cm}
	\vspace{-1cm}
   	\psfrag{u}{\boldmath\footnotesize ${\rm u}$}
   	\psfrag{d}{\boldmath\footnotesize ${\rm d}$}
   	\psfrag{ub}{\boldmath\footnotesize $\bar{\rm u}$}
       	\psfrag{db}{\boldmath\footnotesize $\bar{\rm d}$}
   	\mbox{\epsfig{figure=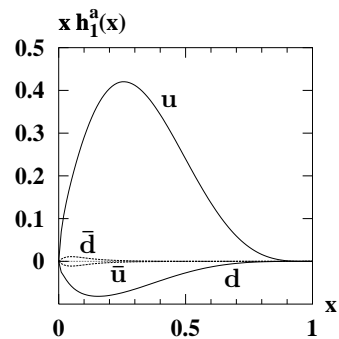,width=5cm,height=5cm}}
   	\caption{\footnotesize\sl
     	  The chiral quark-soliton model prediction for the proton
          $x h_1^a(x)$ vs. $x$ at the scale $Q^2=4\,{\rm GeV}^2$.}
	\vspace{-0.5cm}
	\end{wrapfigure}
%
%
%
%
When applying the DELPHI result eq.(\ref{apower}) to explain HERMES data 
a weak scale dependence of $\la H_1^\perp\ra/\la D_1\ra$ is assumed. 
In ref.\cite{Efremov:2001cz} -- taking the chiral quark soliton model
prediction for $h_1^a(x)$ -- both $H_1^\perp(z) / D_1(z)$ and
$\la H_1^\perp\ra/\la D_1\ra$ have been extracted from 
HERMES and SMC data \cite{hermes-pi0,hermes,bravardis99}.
The value for $\la H_1^\perp\ra/\la D_1\ra$ obtained from that 
analysis is very close to the DELPHI result eq.(\ref{apower}),
but of course model dependent. 
	(The theoretical uncertainty of $h_1^a(x)$ from chiral 
	quark soliton model is around $(10-20)\%$.)
This indicates a weak scale dependence of the analyzing power
and supports the above assumption.
Here we take $\la H_1^\perp\ra/\la D_1\ra = (12.5\pm 1.4)\%$, i.e. the DELPHI
result, eq.(\ref{apower}), with positive sign for which the analysis of 
ref.\cite{Efremov:2001cz} gave evidence for.

\paragraph{\boldmath{$h_1^a(x)$} in nucleon.} 
For the transversity distribution function $h^a_1(x)$ we take predictions 
from the chiral quark-soliton model ($\chi$QSM) \cite{h1-model}.
The $\chi$QSM is a relativistic quantum field-theoretical model with 
explicit quark and antiquark degrees of freedom. This allows to identify 
unambiguously quark as well as antiquark nucleon distribution functions.
The $\chi$QSM has been derived from the instanton model of the QCD
vacuum \cite{DPreview}.
Due to the field-theoretical nature of the $\chi$QSM,
the quark and antiquark distribution functions computed in the model 
satisfy all general QCD requirements (positivity, sum rules, inequalities)
\cite{DPPPW96}.
The model results for the known distribution functions 
-- $f_1^q(x)$, $f_1^{\bar q}(x)$ and $g_1^q(x)$  -- agree within
(10 - 30)$\%$ with phenomenological parameterizations \cite{f1g1-model}. 
This encourages confidence in the model predictions for $h_1^a(x)$.
Fig.1 shows the model results for the proton transversity distribution,
$h_1^{a/p}(x)$ with $a=u,\,\bar u,\,d,\,\bar d$, at $Q^2=4\,{\rm GeV}^2$.

\paragraph{\boldmath$h_1^s(x)$ and $h_1^{\bar s}(x)$ in nucleon.}
We assume strange transversity distributions to be zero
\be\label{h1-strange}
	h_1^s(x)\simeq 0\;,\;\;\; h_1^{\bar s}(x)\simeq 0 \;.\ee
This is supported by calculations of the tensor charge in the SU(3) 
version of the $\chi$QSM \cite{Kim:1996vk}
\ba\label{tensor-charges}
	g_T^s := \int\limits_0^1\!\!\di x \,(h_1^s-h_1^{\bar s})(x)
	      = - 0.008    \;\;\; \mbox{vs.} \;\;\;
	g_T^u =   1.12  \;,\;\;\;
	g_T^d = - 0.42   \ea
at a low scale of $\mu\simeq 0.6\,{\rm GeV}$. 
 (These numbers should be confronted with the realistic $\chi$QSM 
  results for the axial charge
  $g_A^u =  0.902$,
  $g_A^d = -0.478$,
  $g_A^s = -0.054$ \cite{Blotz:1996}.)

The result, eq.(\ref{tensor-charges}) does not necessarily mean that 
$h_1^s(x)$ and $h_1^{\bar s}(x)$ are small {\sl per se}. 
But it makes plausible the assumption, eq.(\ref{h1-strange}), in the sense 
that the strange quark transversity distribution in the nucleon 
can be neglected with respect to the light quark ones.

%
%
	\begin{wrapfigure}{R!}{5.2cm}
	\vspace{-0.5cm}
   	\mbox{\epsfig{figure=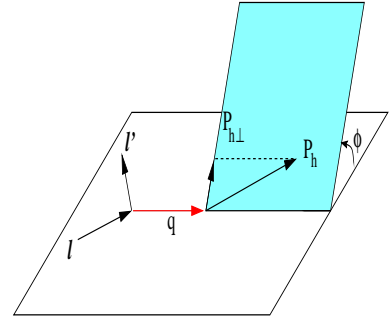,width=5cm,height=4.5cm}}
   	\caption{\footnotesize\sl
    	Kinematics of the process $lD\rightarrow l'h X$ in the lab frame.}
	\vspace{-2cm}
	\end{wrapfigure}
%
%
\paragraph{\boldmath{$h_L^a(x)$} in nucleon.}
The ``twist-3'' distribution function $h_L^a(x)$ can be decomposed as
\cite{Mulders:1996dh}
\be\label{hL-def}
	h_L^a(x)= 2x \int\limits_x^1\!\!\di x'\,\frac{h_1^a(x')}{{x'}^2}
	+ \tilde{h}^a_L(x) \; , \ee
where $\tilde{h}^a_L(x)$ is a pure interaction dependent twist-3 contribution.
According to calculations performed in the instanton model of the QCD vacuum
the contribution of $\tilde{h}^a_L(x)$ in eq.(\ref{hL-def}) is negligible 
\cite{Diakonov:1996qy,Dressler:2000hc}. So when using the $\chi$QSM 
predictions for $h_1^a(x)$ we consequently use the eq.(\ref{hL-def}) 
with $\tilde{h}^a_L(x)\simeq 0$.

\section{Azimuthal asymmetries from deuteron}

In the HERMES experiment the cross sections $\sigma_D^\pm$ for the process 
$lD^\pm\rightarrow l'h X$, see fig.2, will be measured in dependence of 
the azimuthal angle $\phi$ between lepton scattering plane and 
the plane defined by momentum ${\bf q}$ of virtual photon 
and momentum ${\bf P}_{\!h}$ of produced hadron.
($^\pm$ denotes the polarization of the deuteron target, 
$^+$ means polarization opposite to the beam direction.)

Let $P$ be the momentum of the target proton,
$l$ ($l'$) the momentum of the incoming (outgoing) lepton.
The relevant kinematical variables are
center of mass energy square $s:=(P+l)^2$, 
four momentum transfer $q:= l-l'$ with $Q^2:= - q^2$,  
invariant mass of the photon-proton system $W^2:= (P+q)^2$, 
and $x$, $y$ and $z$ defined by
\be\label{notation-1}
	x := \frac{Q^2}{2Pq}	\;,\;\;\;
        y := \frac{2Pq}{s}	\;,\;\;\; 
        z := \frac{PP_h}{Pq\;} 	\;;\;\;\; 
	\cos\theta_\gamma := 1-\frac{2\Mn^2x(1-y)}{s y} \;,  \ee
with $\theta_\gamma$ denoting the angle between target spin and 
direction of motion of the virtual photon.
The observables measured at HERMES are the azimuthal asymmetries 
$A_{UL,D}^{\sin\phi}(x,z,h)$ and $A_{UL,D}^{\sin2\phi}(x,z,h)$ in 
SIDIS electro-production of the hadron $h$. The subscript $_U$ reminds 
on the unpolarized beam, and $_L$ reminds on the longitudinally 
(with respect to the beam direction) polarized deuterium (D) target. 
The azimuthal asymmetries are defined as
\be\label{expl-1}
A_{UL,D}^{W(\phi)}(x,z,h) = 
	\frac{\displaystyle
	   \int\!\!\di y\,\di\phi\,W(\phi)\,\left(
\frac{1}{S^+}\,\frac{\di^4\sigma_D^+}{\di x\,\di y\,\di z \di\phi}-
\frac{1}{S^-}\,\frac{\di^4\sigma_D^+}{\di x\,\di y\,\di z \di\phi}\right)}
     	{\;\;\;\;\;\;\;\displaystyle
	   \frac{1}{2}\int\!\!\di y\,\di\phi\,\left(
   \frac{\di^4\sigma_D^+}{\di x\,\di y\,\di z\di\phi}+
   \frac{\di^4\sigma_D^-}{\di x\,\di y\,\di z\di\phi}\right)}\;\;,\ee
where $W(\phi)=\sin\phi$ or $\sin2\phi$ and 
$S^\pm$ denotes the deuteron spin. For our purposes the 
deuteron cross sections can be sufficiently well approximated by 
\be\label{sigma-d-approx}
	\sigma_D^\pm = \sigma_p^\pm+\sigma_n^\pm\;.\ee
(We do not consider corrections due to deuteron D-state admixture 
which are smaller than other expected experimental and theoretical errors.) 
The proton and neutron semi-inclusive cross sections 
$\sigma^{p\pm}$ and $\sigma^{n\pm}$ eq.(\ref{sigma-d-approx}) have been 
computed in ref.\cite{Mulders:1996dh} at tree-level up to order $1/Q$. 
Using the results of ref.\cite{Mulders:1996dh}
(see ref.\cite{Efremov:2001cz} for an explicit derivation) we obtain
\be\label{AUL-sinPhi}
  	A_{UL,D}^{\sin\phi}(x,z,h) = B_h \, \Biggl(
	P_{\! L}(x)\;
	\frac{\sum_a^h e_a^2\, x h_L^{a/D}(x)\,H_1^{\perp a}(z)}
	     {\sum_{a'}^h e_{a'}^2\, f_1^{a'/D}(x)\,D_1^{a'}(z)\,}
      +	P_{\! 1}(x)\;
	\frac{\sum_a^h e_a^2\, h_1^{a/D}(x)\,H_1^{\perp a}(z)}
	     {\sum_{a'}^h e_{a'}^2\,f_1^{a'/D}(x)\,D_1^{a'}(z)\,}\Biggr)
	\,.\ee
Here, e.g. $h_1^{u/D}(x)=(h_1^{u/p}+h_1^{u/n})(x)=(h_1^u+h_1^d)(x)$,
where as usual $h_1^a(x)\equiv h_1^{a/p}(x)$.  $B_h$ and the 
$x$-dependent prefactors $P_{\!L}(x)$, $P_{\!1}(x)$ are defined by
\ba\label{A-prefactors-def}
	B_h &=& \frac{1}
	{\la z\ra\sqrt{1+\la z^2\ra\la\Pnperp^2\ra/\la\Phperp^2\ra}\;}
	\nonumber\\
	P_{\! L}(x) &=& 
	\frac{\int\!\di y\,4(2-y)\,\sqrt{1-y}\;\cos\theta_\gamma\Mn/Q^5}
	     {\int\!\di y\,(1+(1-y)^2)\,/\,Q^4} \nonumber\\
	P_{\! 1}(x) &=& - \;
	\frac{\int\!\di y\,2(1-y)\,\sin\theta_\gamma/Q^4}
             {\int\!\di y\,(1+(1-y)^2)\,/Q^4}   \;.\ea
The distribution of transverse momenta has been assumed to be Gaussian
which is supported by data \cite{hermes-pi0,hermes}.
$\la\Pnperp^2\ra$ and $\la\Phperp^2\ra$ denote the average transverse 
momentum square of struck quark from the target and of the produced hadron,
respectively. Explicit expression for $A_{UL}^{\sin 2\phi}$ can be found in 
ref.\cite{Efremov:2000za}, which however must be corrected by adding an 
overall sign.

When integrating over $y$ in eq.(\ref{A-prefactors-def})
(and over $z$ and $x$ in the following) one has to consider experimental cuts.
Thereby we neglect the implicit dependence of distribution and fragmentation 
functions on $y$ through the scale $Q^2=xys$, and evaluate them instead 
at $Q^2=4\,{\rm GeV}^2$, typical scale in the HERMES experiment. 
Most cuts used in the data selection are the same as in the 
proton target experiment \cite{hermes-pi0,hermes}
\be\label{exp-cuts}
	1\,{\rm GeV}^2 < Q^2 < 15\,{\rm GeV}^2  , \;\;\;
 	2\,{\rm GeV} < W 			, \;\;\;
	0.2 < y < 0.85 			      \,, \;\;\;
	0.023 < x < 0.4 \ee
and $0.2 <  z  < 0.7$  with $\la z\ra = 0.41$.
Only the cuts for the momentum of the produced hadron 
$2\,{\rm GeV} < |{\bf P}_{\!\!h}| < 15\,{\rm GeV}$ changed
(to be compared with proton target experiments \cite{hermes-pi0,hermes}: 
$4.5\,{\rm GeV} < |{\bf P}_{\!\!h}| < 13.5\,{\rm GeV}$).
The only quantity relevant for our calculation
and possibly altered by these changes is $\la\Phperp^2\ra$.
We assume this change to be marginal -- since upper and lower cut
have been enlarged 'symmetrically' -- and use the value from
refs.\cite{hermes-pi0,hermes}.

\subsection{Pion production}
Due to charge conjugation and isospin symmetry the following relations hold 
\ba\label{eq-favoured-frag}
	D_1^\pi:=
	D_1^{ u/\pi^+}\!  = D_1^{\bar d/\pi^+}\! = 
	D_1^{ d/\pi^-}\!  = D_1^{\bar u/\pi^-}\! =
	2D_1^{ u/\pi^0}\!  = 2D_1^{\bar u/\pi^0}\! =	
	2D_1^{ d/\pi^0}\!\;= 2D_1^{\bar d/\pi^0}\;,\ea
where the arguments $z$ are omitted for brevity.
Analog relations are assumed for $H_1^{\perp\,\pi}$. 
Other ``unfavoured'' fragmentation functions are neglected\footnote{
	In ref.\cite{Ma:2001ie} the effect of unfavoured 
	fragmentation has been studied. The authors conclude that
	'favoured fragmentation approximation' works very well, possibly 
	except for $A_{UL}(\pi^-)$ from a {\sl proton} target.}. 
Then $H_1^{\perp\,\pi}(z)$ and $D_1^\pi(z)$ factorize out 
in eq.(\ref{AUL-sinPhi}) such that
\be\label{AUL-sinPhi-pion}
  	A_{UL,D}^{\sin\phi}(x,z,\pi) = B_\pi \, 
	\frac{H_1^{\perp\,\pi}(z)}{D_1^\pi(z)}\,
	\Biggl(
	P_{\! L}(x)\;
	\frac{\sum_a^\pi e_a^2\, x h_L^{a/D}(x)}
	     {\sum_{a'}^\pi e_{a'}^2\, f_1^{a'/D}(x)\,}
      +	P_{\! 1}(x)\;
	\frac{\sum_a^\pi e_a^2\, h_1^{a/D}(x)}
	     {\sum_{a'}^\pi e_{a'}^2\,f_1^{a'/D}(x)}\Biggr) \,.\ee
Here the summation $\sum_a^\pi$ over those flavours is implied which
contribute to the favoured fragmentation of the pion $\pi$.
So deuteron azimuthal asymmetries in SIDIS pion production are given
(symbolically) by
\be\label{AUL-pion}
	A_{UL,D}^{\sin\phi}(\pi^+) \propto
	\frac{\;4 h_1^{u+d}+h_1^{\bar u+\bar d}\;}
             {\;4 f_1^{u+d}+f_1^{\bar u+\bar d}\;} \;,\;\;
	A_{UL,D}^{\sin\phi}(\pi^0) \propto
	\frac{\;h_1^{u+d}+h_1^{\bar u+\bar d}\;}
             {\;f_1^{u+d}+f_1^{\bar u+\bar d}\;}   \;,\;\;
	A_{UL,D}^{\sin\phi}(\pi^-) \propto 
	\frac{\;h_1^{u+d}+4h_1^{\bar u+\bar d}\;}
             {\;f_1^{u+d}+4f_1^{\bar u+\bar d}\;}   \ee
where $h_1^{u+d}$ is an abbreviation for 
$P_L(h_L^u+h_L^d)(x)+P_1(h_1^u+h_1^d)(x)$ and
$f_1^{u+d}$ for $(f_1^u+f_1^d)(x)$, etc.
Since the $\chi$QSM predicts $(h_1^{\bar u}+h_1^{\bar d})(x)\simeq 0$
\cite{h1-model}, we see from the symbolic eq.(\ref{AUL-pion}) that
the only differences between the asymmetries for different pions are
different weights of the unpolarized antiquark distributions in the
denominator. As $(f_1^u+f_1^d)(x) \gg (f_1^{\bar u}+f_1^{\bar d})(x) > 0$, 
we see that 
\be\label{AUL-pion-2}
	A_{UL,D}^{\sin\phi}(\pi^+)\simgeq 
	A_{UL,D}^{\sin\phi}(\pi^0)\simgeq 
	A_{UL,D}^{\sin\phi}(\pi^-) \; . \ee
Averaging over $z$ in eq.(\ref{AUL-sinPhi-pion})
(numerator and denominator separately), using the {\sl central value}
in eq.(\ref{apower}) for $\la H_1^{\perp\,\pi}\ra / \la D_1^\pi\ra$
and parameterization of ref.\cite{GRV} for $f_1^a(x)$
we obtain the results for $A_{UL,D}^{\sin\phi}(x,\pi)$ shown in fig.3a.
For  $A_{UL,D}^{\sin\phi}(x,\pi^+)$ the statistical error of HERMES data
is estimated.\footnote{The statistical error of 
	$A_{UL,D}^{\sin\phi}(x,\pi^+)$ is estimated by dividing 
	the statistical error of $A_{UL,p}^{\sin\phi}(x,\pi^+)$,
	\cite{hermes}, by $\sqrt{N}$, which considers the roughly
	$N\simeq 3$ times larger statistics of the deuteron target 
	experiment as compared to the proton target experiments
	\cite{www-desy}.}
The small differences between azimuthal symmetries for different pions from 
the deuteron target will be difficult to observe.

We remark that in eq.(\ref{AUL-sinPhi}) the contribution to 
$A_{UL,D}^{\sin\phi}$ containing $h_L^a(x)$ is ``twist-3'' and the 
contribution containing $h_1^a(x)$ is ``twist-2''. 
The ``twist-2'' contribution enters the asymmetry 
with the factor $\sin\theta_\gamma\sim \Mn/Q$, 
see eqs.(\ref{notation-1},\ref{A-prefactors-def}).
So ``twist-2'' and ``twist-3'' are equally power suppressed.
For HERMES kinematics the ``twist-3'' contribution is roughly factor three
larger than the ``twist-2'' contribution and of opposite sign.
However, for larger values of $x>0.4$ the latter becomes dominant, see
Erratum of ref.\cite{Efremov:2001cz}.

For completeness also the $A_{UL,D}^{\sin2\phi}(x,\pi)$ asymmetries are 
shown in fig.3a. 
%
%
\begin{figure}[t!]\label{fig6-predictions}
\begin{tabular}{cccc}
	\hspace{-1cm} &
	\includegraphics[width=7.5cm,height=7.5cm]{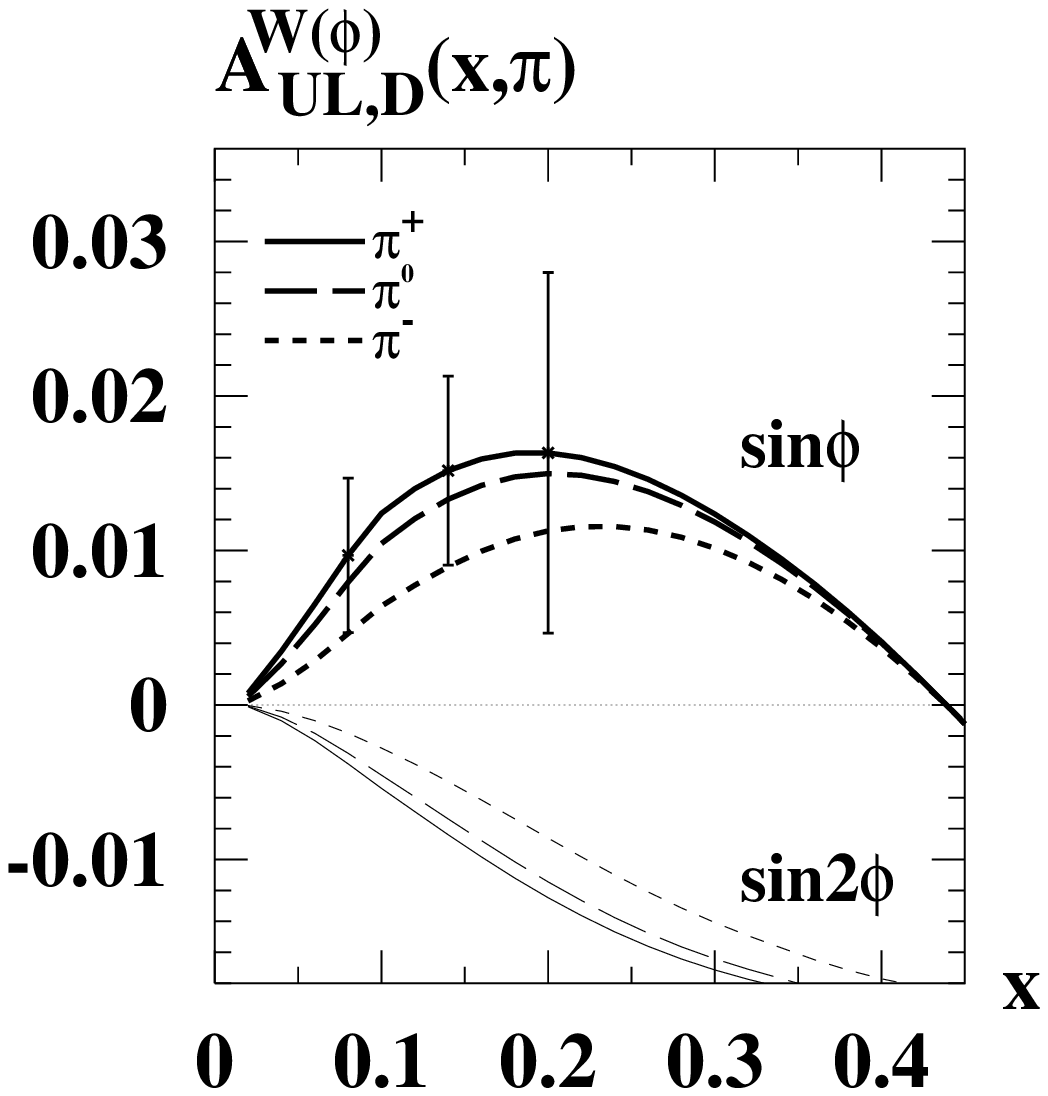} &
	\hspace{-1cm} &
	\includegraphics[width=7.5cm,height=7.5cm]{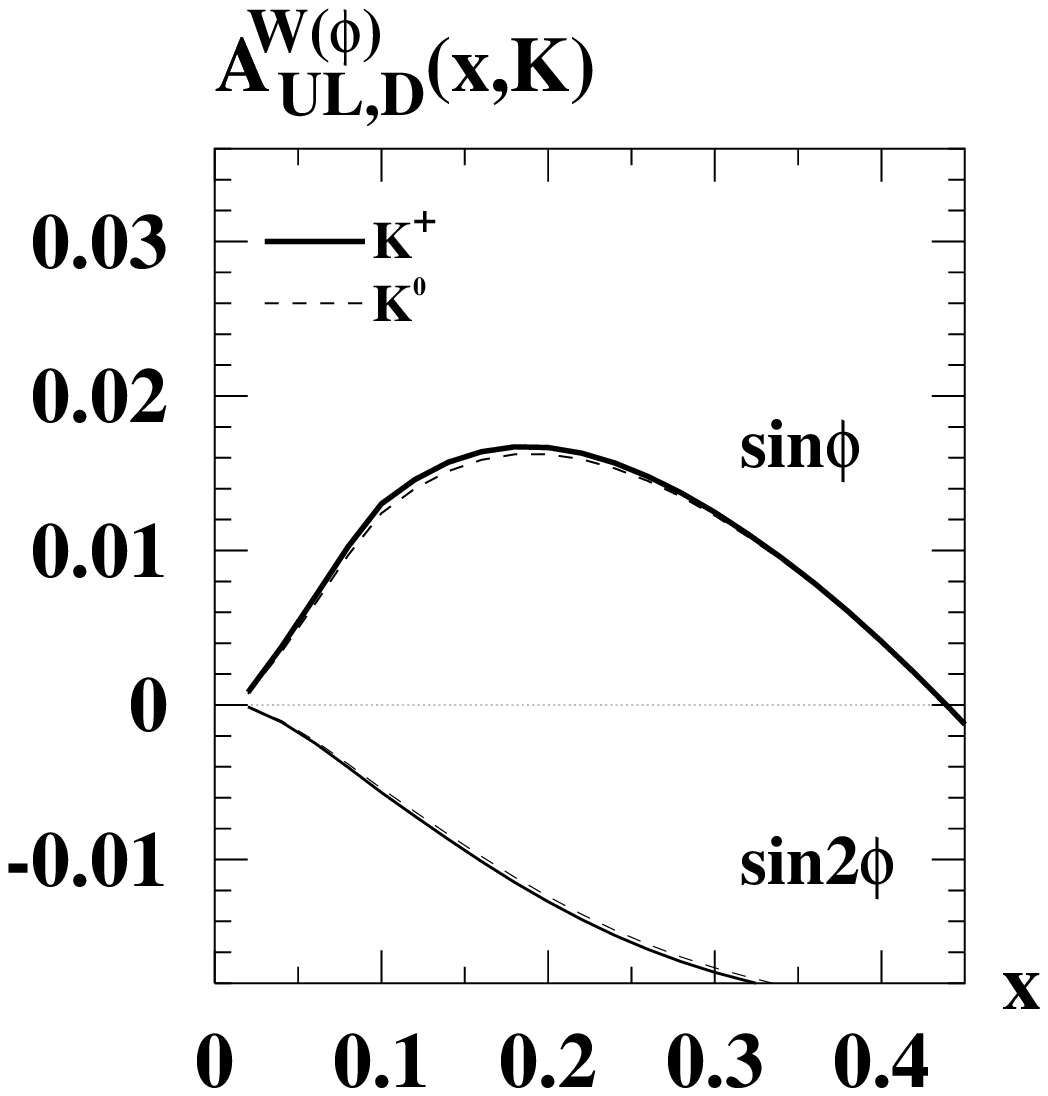}\cr
	\hspace{-1cm} & {\bf a} & \hspace{-1cm} & {\bf b} 
\end{tabular}
	\caption{\footnotesize\sl
	Predictions for azimuthal asymmetries $A_{UL,D}^{W(\phi)}(x,h)$ 
	vs. $x$ from a longitudinally polarized deuteron target for 
	HERMES kinematics.
	{\bf a.} Pions. 
	The "data points" do not anticipate the experiment but correspond 
	merely to a simple estimate of the expected error bars (see text).
	{\bf b.} Kaons, based on assumption eq.(\ref{apower-kaon}).
	The results refer to the central value of the analyzing power
	$\la H_1^\perp\ra/\la D_1\ra=(12.5\pm 1.4)\%$, eq.(\ref{apower}).}
\end{figure}
%
%

\subsection{Kaon production}

The RICH detector of the HERMES experiment is capable to detect kaons. 
For kaons
\be\label{fragment-kaon}
	D_1^K :=	
	D_1^{u/K^+} =  D_1^{d/K^0} = D_1^{\bar d/\bar{K}^0} = D_1^{\bar u/K^-} 
	\:\stackrel{\rm SU(3)}{\simeq}\:
	D_1^{\bar s/K^+} =  D_1^{\bar s/K^0} = D_1^{s/\bar{K}^0} = D_1^{s/K^-}
 	\; . \ee
Analog relations are assumed for $H_1^{\perp K}$.
The exact relations in eq.(\ref{fragment-kaon}) follow from charge 
conjugation and isospin symmetry. The approximate relation follows 
from SU(3) flavour symmetry.\footnote{
	There might be considerable corrections to the approximate 
	SU(3) flavour symmetry relation in eq.(\ref{fragment-kaon}).
	But they have no practical consequences. 
	As for $H_1^{\perp K}$ they do not contribute due 
	to eq.(\ref{h1-strange}). As for $D_1^K$ we have, e.g. for 
	$A_{UL}(K^+)$, in the denominator
	$e_u^2f_1^{u/D}(x)=\frac{4}{9}(f_1^u+f_1^d)(x)\gg 
	 e_s^2f_1^{\bar s/D}(x)=\frac{2}{9}f_1^{\bar s}(x)$.}
As we did for pions, we neglect unfavoured fragmentation into kaons.
So we obtain 
\be\label{AUL-sinPhi-kaon}
  	A_{UL,D}^{\sin\phi}(x,z,K) = B_K \, 
	\frac{H_1^{\perp\,K}(z)}{D_1^K(z)}\,
	\Biggl(
	P_{\! L}(x)\;
	\frac{\sum_a^K e_a^2\, x h_L^{a/D}(x)}
	     {\sum_{a'}^K e_{a'}^2\, f_1^{a'/D}(x)\,}
      +	P_{\! 1}(x)\;
	\frac{\sum_a^K e_a^2\, h_1^{a/D}(x)}
	     {\sum_{a'}^K e_{a'}^2\,f_1^{a'/D}(x)}\Biggr) \,.\ee
We assume that $\la{\bf P}^2_{\!\perp\,K}\ra$ and $\la z\ra$,
which enter the factor $B_K$ eq.(\ref{A-prefactors-def}), 
are the same as for pions. The summation $\sum_a^K$ goes over 
'favoured flavours', i.e. (symbolically)
\be\label{AUL-kaon}
A_{UL,D}^{\sin\phi}(K^+) \propto \frac{\;4h_1^{u+d}}
                                      {\;4f_1^{u+d}+2f_1^{\bar s}\;}\;,\;\;
A_{UL,D}^{\sin\phi}(K^0) \propto \frac{\; h_1^{u+d}}
                                      {\; f_1^{u+d}+2 f_1^{\bar s}\;}\;,\;\;
A_{UL,D}^{\sin\phi}(\bar{K}^0) \simeq 
A_{UL,D}^{\sin\phi}(K^-)       \simeq 0 \;.\ee
Recall that 
$(h_1^{\bar u}+h_1^{\bar d})(x)\simeq h_1^s(x) \simeq h_1^{\bar s}(x)\simeq 0$
according to predictions from $\chi$QSM.

After averaging over $z$ in eq.(\ref{AUL-sinPhi-kaon}) we obtain
$A_{UL,D}^{\sin\phi}(x,K)\propto\la H_1^{\perp\,K}\ra/\la D_1^K\ra$.
How large is the analyzing power for kaons?
We know that the unpolarized kaon fragmentation function $D_1^K(z)$ is 
roughly five times smaller than the unpolarized pion fragmentation 
function $D_1^\pi(z)$ \cite{Binnewies:1994ju}.
Is also $H_1^{\perp\,K}(z)$ five times smaller than $H_1^{\perp\,\pi}(z)$?
If we assume this, i.e. if
\be\label{apower-kaon}
	\frac{\la H_1^{\perp\,  K}\ra}{\la D_1^K  \ra} \simeq
	\frac{\la H_1^{\perp\,\pi}\ra}{\la D_1^\pi\ra} \ee
holds, we obtain -- with the central value of $\la H_1^\perp\ra/\la D_1\ra$ 
in eq.(\ref{apower}) -- azimuthal asymmetries for $K^+$ and $K^0$ 
as large as for pions, see fig.3b. 
	(Keep in mind the different normalization for $H_1^{\perp}$ 
	used here.)
Fig.3b shows also $A_{UL,D}^{\sin2\phi}(x,K)$
obtained under the assumption eq.(\ref{apower-kaon}).

HERMES data will answer the question whether the assumption 
eq.(\ref{apower-kaon}) is reasonable.
In chiral limit $D_1^\pi=D_1^K$ and $H_1^{\perp\pi}=H_1^{\perp K}$
and the relation eq.(\ref{apower-kaon}) is exact.
In nature the kaon is 'far more off chiral limit' than the pion,
indeed $D_1^\pi\gg D_1^K$ \cite{Binnewies:1994ju}. 
In a sense the assumption eq.(\ref{apower-kaon}) formulates the naive
expectation that  the 'way off chiral limit to real world' proceeds 
analogously for spin-dependent quantities, $H_1^\perp$,
and for quantities containing no spin information, $D_1$.

\subsection{Comparison to \boldmath $A_{UL}$ from proton target}
\label{sect-comparison}

$A_{UL,D}^{W(\phi)}(x,\pi^+)$ will be roughly half the magnitude of 
$A_{UL,p}^{W(\phi)}(x,\pi^+)$ which was computed in our approach 
and confronted with HERMES data \cite{hermes} in
ref.\cite{Efremov:2001cz}.
However, the deuteron data will have a smaller statistical error 
due to more statistics. 
So $A_{UL,D}^{\sin\phi}(x)$ for pions and -- upon validity of assumption 
eq.(\ref{apower-kaon}) -- $K^+$ and $K^0$ will be clearly seen in the 
HERMES experiment and perhaps also $A_{UL,D}^{\sin2\phi}(x,h)$. 

In table 1, finally, we present the totally integrated azimuthal
asymmetries $A_{UL}^{\sin\phi}(h)$ for pions from proton target
 -- from ref.\cite{Efremov:2001cz} -- 
and for pions and kaons from deuteron target --  computed here.
HERMES data on  $A_{UL,p}^{\sin\phi}(\pi)$ \cite{hermes-pi0,hermes}
is shown in table 1 for comparison.
The fields with ``?'' will be filled by HERMES data in near future. 

\section{Conclusions}

The approach based on experimental information from DELPHI on
$H_1^\perp$ \cite{todd} and on theoretical predictions from the 
chiral quark soliton model for $h_1^a(x)$ \cite{h1-model} has been shown 
\cite{Efremov:2001cz} to describe well HERMES and SMC data on azimuthal 
asymmetries from a polarized proton target 
\cite{hermes-pi0,hermes,bravardis99}.
Here we computed azimuthal asymmetries in pion and kaon production from 
a {\sl longitudinally polarized deuteron} target for HERMES kinematics.

Our approach predicts azimuthal asymmetries $A_{UL,D}^{W(\phi)}$
comparably large for all pions and roughly half the magnitude of
$A_{UL,p}^{W(\phi)}(\pi^+)$ measured at HERMES \cite{hermes-pi0,hermes}.

Under the assumption that the kaon analyzing power 
$\la H_1^{\perp K}\ra / \la D_1^K\ra$ is as large as the pion
analyzing power $\la H_1^{\perp\pi}\ra / \la D_1^\pi\ra$
we predicted also azimuthal asymmetries for kaons.
If the assumption holds, HERMES will observe $A_{UL,D}^{\sin\phi}(K)$
for $K^+$ and $K^0$ as large as for pions. 
The asymmetries for $\bar K^0$ and $K^-$ are zero in our approach.
It will be exciting to see whether HERMES data will confirm the assumption 
$\la H_1^{\perp K}\ra / \la D_1^K\ra
\:\stackrel{!?}{\simeq}\: \la H_1^{\perp\pi}\ra / \la D_1^\pi\ra$.

	It will be very interesting to study HERMES data on $z$ dependence:
        $A_{UL,D}^{\sin\phi}(z,h)$. 
	The HERMES data on $z$-dependence of azimuthal asymmetries from 
	{\sl proton} target \cite{hermes-pi0,hermes} has been shown 
	\cite{Efremov:2001cz} to be compatible with 
	$H_1^{\perp\pi}(z) / D_1^\pi(z)=a\,z\,$  for  $\,0.2 < z < 0.7$ 
	with a constant $a=(0.33 \pm 0.06 \pm 0.04)$
	(statistical and systematical error of 
	the data \cite{hermes-pi0,hermes}).
	This result has a further uncertainty of $(10-20)\%$ due to
	model dependence.
	Based on this observation we could have predicted here
        $A_{UL,D}^{\sin\phi}(z,h) = c_h\, z$ with $c_h$ some constant
	depending on the particular hadron. It will be exciting to see
	whether HERMES deuterium data will also exhibit a (rough) linear 
	dependence on $z$, or whether it will allow to make a more
	sophisticated parameterization than 
	$H_1^{\perp\pi}(z) / D_1^\pi(z) \propto z$ concluded in
	ref.\cite{Efremov:2001cz}.

\medskip
{\footnotesize
We would like to thank D.~Urbano and M.~V.~Polyakov for fruitful discussions, 
B.~Dressler for the evolution code, and C.~Schill from HERMES 
Collaboration for clarifying questions on experimental cuts.
This work has partly been performed under the contract  
HPRN-CT-2000-00130 of the European Commission.}

%
%
{\footnotesize\begin{table}
\vspace{0.2cm}
\begin{center}
\begin{tabular}{|r||c|c|}  \hline &&\\
\athree{Asymmetry}
       {$A_{UL}^{\sin\phi}(h)$}
       {} & 
\athree{DELPHI + $\chi$QSM for}
       {central value of eq.(\ref{apower}) in $\%$}
       {} &  
\athree{HERMES}
       {$\pm$ stat $\pm$ syst in $\%$}
       {$\phantom{\mbox{central value of eq.(\ref{apower}) in \%}}$} \\
\cline{1-3}\hline\hline &&\\
\athree{proton: \hspace{-0.2cm}}{}{}
\athree{$\pi^+$}{$\pi^0$}{$\pi^-$} & 
\athree{$\phantom{-}2.1$}
       {$\phantom{-}1.5$}
       {$         - 0.3$} & 
\athree{$\phantom{-}2.2 \pm 0.5\pm 0.3$}
       {$\phantom{-}1.9 \pm 0.7\pm 0.3$}
       {$         - 0.2 \pm 0.6\pm 0.4$}\\
&&\\
\cline{1-3}&&\\
\athree{deuteron:\hspace{-0.2cm}}{}{}
\athree{$\pi^+$}{$\pi^0$}{$\pi^-$} & 
\athree{$\phantom{-}1.0$}
       {$\phantom{-}0.9$}
       {$\phantom{-}0.5$} &
\athree{}
       {?}
       {}\\
&& \\
\cline{1-3}&&\\
\athree{deuteron:\hspace{-0.2cm}}{}{}
\athree{$K^+$}{$K^0$}{\hspace{-0.8cm}$\bar{K}^0$, $K^-$} &
\athree{$\phantom{-}1.1$}
       {$\phantom{-}1.0$}
       {$ \,\,\sim 0$} &
\athree{}
       {?}
       {}\\
&& \\ \hline
\end{tabular}
\end{center}\vspace{0.2cm}
\caption{\footnotesize
	Comparison of theoretical numbers and (as far as already measured) 
	experimental data for the totally integrated azimuthal asymmetries 
	$A_{UL}^{\sin\phi}(h)$ observable in SIDIS production of hadron $h$ 
	from {\sl longitudinally polarized proton} and {\sl deuteron} targets,
	respectively.
	{\bf Proton:} HERMES data from refs.\cite{hermes-pi0,hermes}. 
	Theoretical numbers based on DELPHI result eq.(\ref{apower}) and 
	predictions from $\chi$QSM for HERMES kinematics
	from ref.\cite{Efremov:2001cz}.
	{\bf Deuteron:} Predictions from this work for HERMES kinematics.
	HERMES has already finished data taking and is currently analyzing.
	The theoretical numbers have an error due to statistical and
	systematical error of the analyzing power eq.(\ref{apower}) and 
	systematical error of (10-20)$\%$ due to $\chi$QSM.}
\end{table}}
%
%

\end{document}